\documentclass[preprint]{revtex4}
\begin{document}
%\alph{footnote}
\preprint{\bf hep-th/0402224} %\hfill {UCVFC-DF-16-2003}}
\title{ Interacting Particles and Strings in Path and Surface Representations.}
\author{P.J. Arias{${}^{a}$}\footnote{parias@fisica.ciens.ucv.ve}, E.
Fuenmayor{${}^{a}$}\footnote{efuenma@fisica.ciens.ucv.ve} and
Lorenzo Leal{${}^{a}$}\footnote{lleal@fisica.ciens.ucv.ve}}
\address{${}^a$Centro de F\'{\i}sica Te\'orica y Computacional, Facultad de Ciencias, Universidad
Central de Venezuela, AP 47270, Caracas 1041-A, Venezuela.\\}

\begin{abstract}
Non-relativistic charged particles and strings coupled with
abelian gauge fields  are quantized in a geometric representation
that generalizes the Loop Representation. We consider three
models: the string in self-interaction through a Kalb-Ramond field
in four dimensions, the topological interaction of two particles
due to a BF term in $2+1$ dimensions, and the string-particle
interaction mediated by a BF term in $3+1$ dimensions. In the
first case one finds that a consistent "surface-representation"
can be built provided that the coupling constant is quantized. The
geometrical setting that arises corresponds to a generalized
version of the Faraday's lines picture: quantum states are labeled
by the shape of the string, from which emanate ''Faraday`s
surfaces''. In the other models, the topological interaction can
also be described by geometrical means. It is shown that the
open-path (or open-surface) dependence carried by the wave
functional in these models can be eliminated through an unitary
transformation, except by a remaining dependence on the boundary
of the path (or surface). These feature is closely related to the
presence of anomalous statistics in the $2+1$ model, and to a
generalized "anyonic behavior" of the string in the other case.

\end{abstract}
\maketitle
\section{Introduction}
In this paper we study Abelian theories of interacting
non-relativistic point particles and strings, and quantize them in
geometrical representations that generalize the Loop
Representation (LR)\cite{GT}. The interactions are mediated by
Abelian gauge fields, and special attention is devoted to
topological interactions.

 The first model we deal with is that of a string
self-interacting through a Kalb-Ramond field \cite{KR}. This study
is a generalization of that of reference \cite{E}, where charged
non-relativistic point particles in electromagnetic interaction
were quantized within the LR \cite{GT} framework. In \cite{E} it
was found that charge must be quantized in order to the LR
formulation of the model be consistent. This result agrees with
those obtained in previous developments \cite{Corichi1,Corichi2}.

In this work we find that the coupling constant of the string
(let´s say, the Kalb-Ramond ``charge'' of the string) must be
quantized, also, if the geometric representation adapted to the
model is consistent.

Following \cite{E}, we also  consider the LR in the case of
topological interactions. Then, as a second model, we study the
theory of two kinds of non-relativistic point particles, in
$2+1$-dimensions, interacting  through a BF term.  One may couple
minimally the first type of particles to one of the vector fields
and the second type to the other one.  Using this model one
prevents  self-interaction problems that arises in the model of
particles interacting by means of the Chern-Simons field \cite{E}.
This fact leads us to see that in the case of just two particles,
the model precisely  corresponds to the quantum mechanical systen
of two anyons, which has the virtue of being exactly soluble
\cite{Leinas,Wilczek,Wu,Indio}. Furthermore, this "toy-model"
opens the way and gives us the key for  understanding  the more
involved theory of the next section.

This last model consists of a particle and a string interacting
through a BF topological term in $3+1$-dimensions.
 Again,  the particle couples in a natural way with the $1$-form
 ($A_{\mu}$), while the string does it with the $2$-form ($B_{\mu\nu}$). The
quantization can be done in two "dual" geometric frameworks: a
path and a surface representation.

As in references \cite{E,LO}, when the  fields that provide the
interaction have a topological character, the dependence of the
wave-functionals on paths (in general, on the appropriate
geometric objects that enter in the representation, like paths or
surfaces) may be eliminated by means of an unitary transformation.
In that case one obtains a quantum mechanics of particles, or
particles and strings (depending on the model), subjected to a
long range interaction.

As in the particle-field interaction \cite{E}, the coupling of
extended ``matter'' objects to fields presents certain subtleties
regarding its quantization, and so does the appropriate geometric
representation. We shall deal with an extension of the
conventional LR, namely, the surface representation, which was
considered several years ago to study the free-field case
\cite{PIO,Pio}, but has to be adapted to include the
particularities that the coupling with the string demands .

The paper is organized as follows. In section \ref{sec2} we study
a geometric surface representation for the non-relativistic
``charged'' string in  Kalb-Ramond interaction. Section \ref{sec3}
is dedicated to consider the  path representation quantization of
two different species of  non-relativistic point particles
interacting by means of a topological BF term in $2+1$-dimensions.
We devote section \ref{sec4} to the study of the interaction of
non-relativistic point charged particles and strings through a
topological BF term in $3+1$-dimensions. Some discussions and
final remarks are given in the last section.

\section{Non-relativistic string interacting with the Kalb-Ramond field. Surface-representation.}\label{sec2}

The first model we are going to study is described by the action
\begin{eqnarray}\label{2.1}
S=-\frac{1}{12{g^2}}\int H^{\mu\nu\lambda}H_{\mu\nu\lambda}d^{4}x
+ \frac{\alpha}{2}\int dt \int d\sigma
\left[(\dot{z}^i)^2-(z'^i)^2) \right] + \frac{1}{2} \int d^4x
\textit{J}^{\mu\nu}B_{\mu\nu}\nonumber\\
\end{eqnarray}
Although the theory is not Lorentz invariant, we find it
convenient to employ four-space notation to some extent. The
Kalb-Ramond antisymmetric potential and field, $B_{\mu\nu}$ and
$H_{\mu\nu\lambda}$, are related by
$H_{\mu\nu\lambda}=3\partial_{[\mu}B_{\nu\lambda]}=\partial_{\mu}
B_{\nu\lambda}+\partial_{\lambda} B_{\mu\nu}+\partial_{\nu}
B_{\lambda\nu}$. Besides the Kalb-Ramond term (the first one), the
action comprises a contribution corresponding to the free
non-relativistic closed string, whose world sheet spatial
coordinates $z^{i}(t,\sigma)$ are given in terms of the time $t$
and the parameter $\sigma$ along the string (dot and prime
indicate derivatives with respect to $t$ and $\sigma$
respectively). $\alpha$ is the string tension, having units of
$mass^2$ and $g$ is a parameter with units of $mass$ in order to
have a dimensionless coupling constant between the string and the
Kalb-Ramond potential. In the string-field interaction term, the
current $J^{\mu\nu}$ associated to the string is given by
\begin{equation}\label{2.2}
\textit{J}^{\mu\nu}(\vec{x},t)= \phi \int dt \int d\sigma
\left[\dot{z}^\mu z'^\nu -\dot{z}^\nu
z'^\mu\right]\delta^{(4)}(x-z),
\end{equation}
where $\phi$ is the dimensionless coupling constant. This current
will be dynamically conserved. The interaction term can be written
as
\begin{equation}\label{2.3}
S_{int}=\frac{\phi}{2} \int dt \int d\sigma \left[\dot{z}^\mu
z'^\nu-\dot{z}^\nu z'^\mu\right]B_{\mu\nu}(z).
\end{equation}

The generalization of what we are going to study to the case of
more than one string, even with different couplings for each one,
is straightforward. For the sake of simplicity we shall mainly
consider the model with just one string; some remarks about the
general case are given at the end of this section. The action is
invariant under the gauge transformations
\begin{equation}\label{2.4}
\delta B_{\mu\nu}=2\partial_{[\mu}\lambda_{\nu]}=
\partial_{\mu}\lambda_{\nu}-\partial_{\nu}\lambda_{\mu},
\end{equation}
provided the string is closed. We are interested in performing the
Dirac quantization of the theory. To this end, we need the $3+1$
decomposition of the action [we are employing the ``metric"
$\eta_{\mu\nu}=(+,-,-,-)$]
\begin{equation}\label{2.5}
S=\int d^4x\left(-\frac{1}{12g^2}H_{ijk}H_{ijk}+
\frac{1}{4g^2}H_{0ij}H_{0ij}+B_{0i}\textit{J}^{0i}+\frac{1}{2}B_{ij}\textit{J}^{ij}\right)
+\frac{\alpha}{2}\int dt \int d\sigma\left[(\dot{z}^i)^2-(z'^i)^2)
\right],
\end{equation}
so the conjugate momenta associated to the fields, $B_{ij}$, and
string variables, $z^i$, are
\begin{equation}\label{2.6}
\Pi^{ij}=\frac{1}{2g^2}\left(\dot{B}_{ij}+ \partial_j
B_{0i}-\partial_i B_{0j}\right), \quad \quad P_{i}=\alpha
\dot{z}^i +\phi B_{ij}z'^{j}.
\end{equation}
The field variables $B_{i0}$, which have vanishing momenta, are
treated as non-dynamical fields from the very beginning. In fact,
the Hamiltonian results to be
\begin{equation}\label{2.7}
H=\int d^3x \left[{g}^2
\Pi^{ij}\Pi^{ij}+\frac{1}{12g^2}H_{ijk}H_{ijk}\right]+\int d\sigma
\frac{\alpha}{2}\left[\frac{1}{\alpha^2}\left(P_{i}-\phi
B_{ij}(z)z'^{j}\right)^{2}+(z'^{i})^2 \right]+\int d^3x
B_{0i}\chi^i ,
\end{equation}
hence, the role of $B_{i0}$ as Lagrange multipliers enforcing the
constraints
\begin{equation}\label{2.8}
\chi^{i}(x) \equiv - \rho^{i}(x) - 2\partial_j \Pi^{ji}(x))=0,
\end{equation}
with
\begin{equation}\label{2.9}
\rho^{i}(x)\equiv \phi \int d\sigma
z'^i\delta^{(3)}(\vec{x}-\vec{z}),
\end{equation}
becomes evident.

The canonical Poisson brackets are defined as
\begin{equation}\label{2.10}
\left\{z^i(\sigma),P_j(\sigma')\right\}=\delta^i_j\delta(\sigma
-\sigma'),
\end{equation}
\begin{equation}\label{2.11}
\left\{B_{ij}(\vec{x}),\Pi^{kl}(\vec{y})\right\}=\frac12\left(\delta_{i}^{k}\delta_{j}^{l}
-\delta_{i}^{l} \delta_{j}^{k}\right)
\delta^{(3)}(\vec{x}-\vec{y}).
\end{equation}
The remaining Poisson brackets vanish.

The preservation of the constraints given above does not produce
new ones. Furthermore, they result to be first class constraints
that generate time independent gauge transformations on the phase
space of the theory.

The basic observables in the sense of Dirac that can be
constructed from the canonical variables are the generalized
electric and magnetic fields
\begin{equation}\label{2.12}
\Pi^{ij}=\frac{1}{2g^2}H_{0ij}\equiv \frac{1}{2g^2}E^{ij},
\end{equation}
\begin{equation}
\textbf{B}\equiv \frac{1}{3!} \epsilon^{ijk}H_{ijk},
\end{equation}
the position $z^{i}(\sigma)$, and the covariant momentum of the
string
\begin{equation}\label{2.13}
P_{i}-\phi B_{ij}(z)z'^{j}.
\end{equation}

All the physical observables of the theory are built in terms of
these gauge invariant quantities, as can be verified. For
instance, the Hamiltonian, given in equation (\ref{2.7}) fulfils
this requirement.

To quantize, the canonical variables are promoted to operators
obeying the commutators that result from the replacement
$\{\;,\;\}\;\longrightarrow\;-i[\;,\;]$. These operators have to
be realized in a Hilbert space of physical states
$|\Psi\rangle_{Phys}$, that obey the generalized Gauss law
\begin{equation}\label{2.15}
-\left(\rho^{i}(x) + 2\partial_j \Pi^{ji}(x)\right)\mid
\Psi_{Physical}\rangle\approx 0.
\end{equation}

At this point, we introduce a geometric representation adapted to
the present model. It will be an ``open-surfaces representation'',
which is closely related with the LR as formulated by Gambini and
Tr\'{\i}as \cite{GT}, and with an early geometrical formulation of
the pure Kalb-Ramond field, based on closed surfaces
\cite{PIO,Pio}.

Consider the space of piecewise smooth oriented surfaces (for our
purposes) in $R^{3}$. A typical element of this space, let say
$\Sigma$, will be the union of several surfaces, perhaps some of
them being closed. In this space we set up the following
equivalence relation: we identify two $\Sigma 's$ that share the
same "form factor" $T^{ij}(x,\Sigma)$ defined as
\begin{equation}\label{2.16}
T^{ij}(x,\Sigma)=\int
d\Sigma^{ij}_y\,\delta^{(3)}(\vec{x}-\vec{y}).
\end{equation}
with $d\Sigma^{ij}_y=(\frac{\partial y^i }{\partial
s}\frac{\partial y^j}{\partial r} - \frac{\partial y^i}{\partial
r}\frac{\partial y^j}{\partial s})ds dr $, $s,r$ being parameters
for the surface. It is easy to show that this indeed defines an
equivalence relation. Also, observe that two surfaces differing in
the parametrization belong to the same class, since they trivially
have the same form factor.

It is worth noticing that the composition of surfaces, together
with the equivalence relation stated above, define a group product
among the classes. The resulting group is Abelian, since the form
factor of the composition of two $\Sigma 's$ is the sum of their
respective form factors. All these features of the``open surfaces
space", are more or less immediate generalizations of  aspects
already encountered in its one dimensional relative, the Abelian
path space \cite{GT,C,LO,E,LL}.

Now we consider functionals $\Psi(\Sigma)$ depending on classes
$\Sigma$ [we employ the same notation both for the surface and the
class to which it belongs, since from now on all the
surface-dependent objects that will appear are indeed
class-dependent ones]. We introduce the surface derivative
$\delta_{ij}(x)$, that measures the response of $\Psi(\Sigma)$
when an element of surface whose infinitesimal area is
$\sigma_{ij}$ is attached to the argument $\Sigma$ of
$\Psi(\Sigma)$ at the point $x$, up to first order in
$\sigma_{ij}$
\begin{equation}\label{2.17}
\Psi (\delta\Sigma\cdot\Sigma)-\Psi
(\Sigma)=\sigma^{ij}\delta_{ij}(x)\Psi (\Sigma)
\end{equation}
where
\begin{equation}\label{2.18}
\sigma^{ij}=u^iv^j-v^ju^i ,
\end{equation}
is the surface element generated by the infinitesimal vectors
$\vec{u}$ and $\vec{v}$. The surface derivative $\delta_{ij}(x)$
should not be confused with the loop derivative $\Delta_{ij}(x)$.
Unlike the former, the latter acts onto loop-dependent
functionals. Of course, since in $R^{3}$ loop-dependence is a
particular case of surface-dependence (a loop can be seen as the
boundary of an open surface, whenever the manifold be trivial in
the homological sense) , the loop derivative can be taken as the
surface derivative restricted to loop-dependent functionals. In
this sense it can be said that $\delta_{ij}(x)$ ``includes''
$\Delta_{ij}(x)$.

  From $\delta_{ij}(x)$ it is possible to define the closed-surface
derivative $\triangle_{ijk}$ of reference \cite{Pio}, that appends
a small cube of volume $V^{ijk}$ to the argument $\Sigma$ of
$\Psi(\Sigma)$. The relation between both derivatives is
\begin{equation}\label{2.19}
\triangle_{ijk}(x)=\partial_i\delta_{jk}(x)
+\partial_j\delta_{ki}(x) +\partial_k\delta_{ij}(x).
\end{equation}
It should be noticed that the right hand side of the above
equation would vanish if $\delta_{ij}(x)$ be the same as
$\Delta_{ij}(x)$. This is so, since $\Delta_{ij}(x)$ is the curl
of a more basic object: the path-derivative (see equation
\ref{3.19}) \cite{Pio}.

Turning back to the quantization of our model, it can be seen that
the fundamental commutator associated to equation (\ref{2.11}) can
be realized on surface-dependent functionals if one prescribes

\begin{equation}\label{2.20}
\hat{\Pi}^{ij}(\vec{x})\longrightarrow
\frac{1}{2}T^{ij}(\vec{x},\Sigma),
\end{equation}
\begin{equation}\label{2.21}
\hat{B}_{ij}(\vec{x})\longrightarrow 2i\delta_{ij}(\vec{x}),
\end{equation}
since the surface-derivative of the form factor is given by
\begin{equation}\label{2.22}
\delta_{ij}(\vec{x}) T^{kl}(\vec{y},\Sigma)=\frac
12\left(\delta_{i}^{k}\delta_{j}^{l} -\delta_{i}^{l}
\delta_{j}^{k}\right)\delta^{(3)}(\vec{x}-\vec{y}).
\end{equation}

On the other hand, the operators associated to the string can be
realized in a ``shape" representation, i.e., onto functionals
$\Psi [z(\sigma)]$ that depend on the (spatial) coordinates of the
string world sheet

\begin{equation}\label{2.23}
\hat{z}^i(\sigma) \longrightarrow z^i(\sigma), \quad
\hat{P}^i(\sigma)\longrightarrow -i \frac{\delta} {\delta
z^i(\sigma)} .
\end{equation}
Henceforth, the states of the interacting theory can be taken as
functionals $\Psi[\Sigma,z(\sigma)]$ depending both on surfaces
(i.e. the equivalence classes discussed above) and functions
$z(\sigma)$. Among these functionals, we must pick out those that
belong to the kernel of the Gauss constraint (\ref{2.15}), that in
this representation can be written as
\begin{eqnarray}\label{2.24}
&&\left(\;\rho^{i}(\vec{x}) +
2\partial_j\Pi^{ji}(\vec{x})\;\right)
\Psi[\Sigma,z(\sigma)]\approx 0\quad\Longrightarrow\nonumber\\
&&\left(\phi\int_{string} d\sigma
z'^i\delta^{(3)}(\vec{x}-\vec{z}) -\int_{\partial\Sigma}d\sigma
z'^i \delta^{(3)}(\vec{x}-\vec{z})\right)\Psi[\Sigma ,z(\sigma)]
\approx 0,
\end{eqnarray}
where we have used
\begin{equation}\label{2.25}
\partial_j T^{ji}(\vec{x},\Sigma)=-T^i(\vec{x},\partial\Sigma)
=-\int_{\partial\Sigma} dz^i \delta^{(3)}(\vec{x}-\vec{z}),
\end{equation}
with $\partial \Sigma$ being the boundary of $\Sigma$. To solve
this constraint, it is useful to recall which is the geometrical
setting that allows to solve the Gauss constraint in  Maxwell
theory coupled to non-relativistic particles, which was discussed
in reference \cite{E}. There, the appropriate physical space can
be labelled by lines of Faraday: every particle carries a bundle
of lines emanating from or arriving to the particle (depending on
the sign of the particle's charge). This construction is possible
only if charge is quantized, since the number of Faraday lines,
which must be equal to the charge to which they are attached, has
to be an integer. In the present case, we see that if the surface
is such that its boundary coincides with the string, the
constraint (\ref{2.24}) reduces to
\begin{equation}\label{2.26}
\left(\phi-1\right)\,\int_{string}d\sigma z'^j
\delta^{(3)}(\vec{x}-\vec{z})=0,
\end{equation}
and it is satisfied for $\phi =1$. In that case it can be said
that the surface emanates from the string. It could well happen
that, instead, the boundary of the surface had the opposite
orientation of the string. Then, the constraint would demand that
$\phi =-1$, and we say that the surface ends at the string.
Clearly, there is also the possibility that the surface be
composed by several layers, say $n$ of them, that start (or end)
at the string. In this situation, equation (\ref{2.24}) becomes
\begin{equation}\label{2.26a}
\left(\phi-n\right)\,\int_{string}d\sigma z'^j
\delta^{(3)}(\vec{x}-\vec{z})=0,
\end{equation}
and the coupling constant must obey $\phi =n$. The sign of $n$
depends on whether the layers are "incoming" or "outgoing", in the
sense explained above. Finally, it should be remarked that when
$\phi =n$, the surface may consist of the $n$ layers attached to
the string, plus an arbitrary number of closed surfaces, since the
latter  do not contribute to the boundary of the surface that
define the equivalence class $\Sigma$.

Thus, we find that the physical sector of the Hilbert space of the
theory, in the surface-representation, consists of wave
functionals that depend on "surfaces of Faraday" for the
string-Kalb-Ramond system. Notice that in the case of $N$ strings,
carrying different ``charges'' $\phi_{a}$, $a=1,..,N$, each string
must be a source or sink of its own bundle of  $n_{a}=\phi _{a}$
layers (as before, these bundles may be accompanied by closed
pieces of surfaces). This geometrical setting is possible if the
couplings $\phi_{a}$ are quantized, since each individual sheet or
layer carries a unit of Kalb-Ramond electric flux.

To conclude this section, let us write down the Schr\"{o}dinger
equation of the model in the surface-representation
\begin{eqnarray}\label{2.27}
-i\frac{\partial}{\partial t}\Psi[\Sigma
,z(\sigma)]&=&H\Psi[\Sigma,z(\sigma)]\nonumber\\
&=&\frac{1}{2g^2}\int d^3x \left[\textbf{B}^2+\frac{1}{2}
E^{ij}E^{ij}\right]+\nonumber\\
&+&\int d\sigma\frac{\alpha}{2}\left[\frac{-1}{\alpha^2}
\left(\frac{\delta}{\delta z^i}+2\phi\,{\delta}_{ij}
(\vec{z})z'^{j}\right)^{2}+(z'^{i})^2\right]\Psi[\Sigma,z(\sigma)].
\end{eqnarray}
The first term correspond to the free-field contributions to the
energy of the system. In fact, $\textbf{B}^2$ is a kind of
"surface laplacian", while $E^{ij}E^{ij}$ (which indeed contains a
square of Dirac-delta-functions, hence it should be regularized)
may be though as the "position of the surface" squared. The
remaining term correspond to  the string energy, taking into
account the minimal coupling to the Kalb-Ramond field.  It should
be noticed that every term in the right hand side of this
expression respects the geometrical properties of the physical
sector that we have studied in the previous discussion. For
instance, the covariant momentum $-i\left(\frac{\delta}{\delta
z^i}+2\phi\, {\delta}_{ij}(\vec{z})z'^{j}\right)$, which encodes
the field-string interaction, acts onto the wave functionals
$\Psi[\Sigma,z(\sigma)]$ as a generalized Mandelstam derivative
\cite{MDS}: while the functional derivative  with respect to
$z^{i}(\sigma)$ translates (infinitesimally) the string, the
surface derivative evaluated at the string coordinate $\sigma$,
times $\phi$, serves to join the infinitesimally translated string
to the bundle of layers that, otherwise, would remain separated of
the string, breaking gauge invariance.

\section{Toy model: non-relativistic particles interacting through a  BF term in $2+1$ dimensions }\label{sec3}

In this section we shall study the path representation of a BF
term in $2+1$ dimensions coupled with two types of dynamical
particles. This "toy model" already exhibits many of the features
that we shall encounter in  section \ref{sec4}, where we shall
deal with a $3+1$-dimensions topologically interacting
particle-string model.

The action that we shall take is written as

\begin{equation}\label{3.1} S=\frac12\int
\epsilon^{\mu\nu\lambda}B_{\mu}F_{\nu\lambda}d^{3}x + \int dt
\,\left(\frac12 \,m\dot{\vec{r}}^{\,2}+\frac12
\,M\dot{\vec{R}}\right)+\int d^{3}x\left(j^{\mu}A_{\mu} +
J^{\mu}B_{\mu}\right),
\end{equation}
where
$F_{\nu\lambda}=\partial_{\nu}A_{\lambda}-\partial_{\lambda}A_{\nu}$.
We use small and capital letters to distinguish the quantities
related with the two types of particles. The idea we have in mind
is to generalize this model in the next section replacing the
``big'' particles by an extended object (string). The current
$\textit{j}^{\,\mu}$,
 is given by
\begin{equation}\label{3.2}
j^{\mu}(\vec{x})= q v^{\mu}\delta^{(2)}(\vec{x}-\vec{r})=(\rho
(\vec{x}),\vec{j}(\vec{x})),
\end{equation}
where $q$ is the charge of the "type one" particle coupled with
the $1$-form $A_{\mu}$ and $v^{\mu}=(1,\dot{\vec{r}})$ its
velocity. A similar expression holds for $\textit{J}^{\mu}$, with
capital letters replacing small ones. The dimensions of $A_{\mu}$
and $B_{\mu}$ are $length^{-1}$ so $q$ and $Q$ are dimensionless.

It should be observed that the BF term can be decoupled in two
Chern -Simons terms via the field transformation $A_{\mu}\approx
a^1_{\mu}+a^2_{\mu}$, $B_{\mu}\approx a^1_{\mu}-a^2_{\mu}$.
Nevertheless, when the source and particle terms are present the
system does no decouple directly in two particle-Chern-Simons
models.

The $2+1$ decomposition of the action is given by
\begin{eqnarray}\label{3.3}
S&=&\int d^{3}x
\left(\epsilon^{ij}\partial_{i}A_{j}B_{0}+\epsilon^{ij}\partial_{i}B_{j}A_{0}+\dot{A}_{j}\epsilon^{ij}B_{j}
\right)+\int dt \,\left(\frac12 \,m\dot{\vec{r}}^{2}+\frac12
\,M\dot{\vec{R}}^{2}\right)\nonumber\\&+&\int
dt\left(qA_{0}(\vec{r})+qA_i\dot{r}^i\right)+\int dt
\left(QB_{0}(\vec{R})+QB_i\dot{R}^i\right).
\end{eqnarray}

To perform the Dirac quantization of the model we will not take
$A_{0}$, $B_{0}$ as true dynamical variables. Moreover, the
decomposition (\ref{3.5}) shows that $\epsilon^{ij}B_{j}$ is the
conjugated momentum of $A_{i}$, and there is no need to treat
$B_{i}$ as an independent "generalized coordinate" \cite{Jackiw2},
instead we will take $\frac{\partial{\cal
L}}{\partial\dot{A}_i}\equiv\Pi^i=\epsilon^{ij}B_{j}$ as a
definition. Hence, the conjugate momenta associated to the fields
and particles variables are
\begin{eqnarray}\label{3.5}
&&\Pi^{i}=\epsilon^{ij}B_{j},\quad \quad {p_{i}}=m\dot{r}^{i}+q A_{i}(\vec{r}),\\
&&{P_{i}}=M\dot{R}^{i}+Q B_{i}(\vec{r}).
\end{eqnarray}
The Hamiltonian has the form
\begin{eqnarray}\label{3.6}
H&=&\frac{\Big(\vec{P}-q\vec{A}(\vec{r})\Big)^2}{2m}+
\frac{\Big(\vec{P}-Q\vec{B}(\vec{R})\Big)^2}{2M}+\int
d^{2}x\left[A_{0}(x)\chi_{1}(x)+B_{0}(x)\chi_{2}(x)\right]\nonumber\\
&\equiv & H_{0} + \int
d^{2}x\left[A_{0}(x)\chi_{1}(x)+B_{0}(x)\chi_{2}(x)\right],
\end{eqnarray}
where we have defined $\chi_{1}(x)$ and $\chi_{2}(x)$ as
\begin{eqnarray}\label{3.7}
\chi_{1}(x)\equiv -\epsilon^{ij}\partial_{i}B_{j}+
q\delta^{(2)}(\vec{x}-\vec{r})\equiv
\textbf{B}_{\vec{B}}-\rho_{1}(x),\nonumber\\
\chi_{2}(x)\equiv-\epsilon^{ij}\partial_{i}A_{j}
-Q\delta^{(2)}(\vec{x}-\vec{R})\equiv
\textbf{B}_{\vec{A}}-\rho_{2}(x).
\end{eqnarray}
In (\ref{3.6}), the role of $A_{0}(x)$ and $B_{0}(x)$ as Lagrange
multipliers that enforce the secondary constraints $\chi_{1}(x)=0$
and $\chi_{2}(x)=0$ becomes clear. In the last equations, we have
defined the magnetic fields associated to $A_{\mu}$ and $B_{\mu}$
\begin{equation}\label{3.4}
\textbf{B}_{\vec{A}}=-\frac12\epsilon^{ij}F_{ij}=-\epsilon^{ij}\partial_{i}A_{j},\qquad
\textbf{B}_{\vec{B}}=-\epsilon^{ij}\partial_{i}B_{j}.
\end{equation}
It should be recalled that the BF term, being topological, does
not contribute to the energy-momentum tensor.  That is why the
Hamiltonian $H_0$ has the form of that of a collection of two sets
of particles in different external fields. This feature also
appears when dealing with the theory or particles interacting
through a Chern-Simons field \cite{E}.

The canonical commutators are defined as
\begin{eqnarray}\label{3.8}
\left[{r^i},{p_{j}}\right]&=& i\delta^i_j,\nonumber\\
\left[{R^i},{P_j}\right]&=&i\delta^i_j,\nonumber\\
\left[A_i(\vec{x}),
\epsilon^{jk}B_{k}(\vec{y})\right]&=&i\delta^j_i\delta^{(2)}(\vec{x}-\vec{y}).
\end{eqnarray}
 The remaining ones vanish identically. The constraints $\chi_{1}(x)$ and $\chi_{2}(x)$
 ,written in (\ref{3.7}), are readily seen to be of first class.

The Gauge invariant observables (in Dirac`s sense) that can be
constructed from the canonical variables are the generalized
magnetic fields defined in (\ref{2.4}), the positions $\vec{r}$,
$\vec{R}$ and the ``covariant'' momenta $p_i-qA_i(\vec{r})$ and
$P_i-QB_i(\vec{R})$ . As before, all the physical observables of
the theory are built in terms of these gauge invariant quantities,
as can be easily verified. These fundamental observables have to
be realized in a Hilbert space of physical states
$|\psi\rangle_{Phy}$, that obey two generalized Gauss laws (one
for each type of particle) given by
\begin{eqnarray}\label{3.10}
\chi_{1}(x)\mid \psi_{Physical}\rangle &=& -\left({\rho}_1(x)
+\epsilon^{ij}\partial_{i}B_{j}(x)\right)\mid \psi_{Physical}\rangle\approx 0,\nonumber\\
\chi_{2}(x)\mid\psi_{Physical}\rangle &=&-\left({\rho}_2(x) +
\epsilon^{ij}\partial_i A_{j}(x)\right)\mid
\psi_{Physical}\rangle\approx 0.
\end{eqnarray}
 At this point, we introduce a geometric representation adapted
to the present model. It is the Abelian path representation
\cite{GT,E,LL,LO,C}, that can be summarized as follows. Consider
the space of oriented open paths in $R^{2}$. An element $\gamma$
of this space will be the union of several curves, perhaps some of
them being closed. As we did for the surface representation, we
set up an equivalence relation by defining the ``form factor'' of
the curves
\begin{equation}\label{3.11}
T^{i}(x,\Sigma)=\int_{\gamma}
dy^{i}\,\delta^{(2)}(\vec{x}-\vec{y}),
\end{equation}
and  state that two  curves $\gamma$, $\gamma'$ are equivalent if
they share the same form factor. Every equivalence class defines
what we shall call a path. The composition of curves, together
with the equivalence relation defines a group product among
classes of equivalence, i.e., among paths. It can be shown that
this group is Abelian \cite{GT}. Now, let us consider
path-dependent functionals  $\Psi(\gamma)$ [we employ the same
notation for curves and paths]. We introduce  the path derivative
$\delta_i(x)$, that measures the change in $\Psi(\gamma)$ when an
``infinitesimal'' path $u_{\vec{x}}$ is attached to the argument
$\gamma$ of $\Psi(\gamma)$ at the point $x$, up to first order in
the vector $\vec{u}$ associated to the path we have
\begin{equation}\label{3.12}
\Psi(\gamma \cdot u_{\vec{x}}) = \Psi (\gamma) + u^i \delta_i
(\vec{x})\Psi (\gamma).
\end{equation}

One also has a loop derivative \cite{GT} $\Delta_{ij}(\vec{x})$
defined as
\begin{equation}\label{3.13}
\Psi (\sigma \cdot C)= \left(1+
\sigma^{ij}\Delta_{ij}(\vec{x})\right)\, \Psi (C),
\end{equation}
with $C$ being a closed path (a loop) and $\sigma^{ij}$ being the
area enclosed by an infinitesimal loop attached at the spatial
point $\vec{x}$. Thus $\Delta_{ij}(\vec{x})$ measures how the loop
dependent function $\Psi (C)$ changes under a small deformation of
its argument $C$. The loop derivative is readily seen to be the
curl of the path derivative \cite{C},
\begin{equation}\label{3.19}
\Delta_{ij}(\vec{x})= \frac{\partial}{\partial
x^{i}}\delta_{j}(\vec{x})-\frac{\partial}{\partial
x^{j}}\delta_{i}(\vec{x}).
\end{equation}
 The canonical algebra
(\ref{3.8}) can be realized by means of the prescriptions
\begin{eqnarray}\label{3.14}
\hat{A}_{i}(\vec{x})&\longrightarrow
& i\,\delta_i(\vec{x}),\nonumber\\
\hat{\Pi}^{i}(\vec{x})&\longrightarrow & \,
T^{i}(\vec{x},\gamma),\nonumber\\
\hat{r^{i}}\longrightarrow {r^{i}}\,,&&
\hat{p_{j}}\longrightarrow -i \frac{\partial}{\partial {r^{j}}}\\
\hat{R^{i}}\longrightarrow {R^{i}}\,,&& \hat{P_{j}}\longrightarrow
-i \frac{\partial}{\partial {R^{j}}}\nonumber
\end{eqnarray}
These operators act onto wave functionals $\Psi[\gamma, \vec{r},
\vec{R}]$ that depend on the path $\gamma$ and the positions of
both types of particles $\vec{r}$, $\vec{R}$. To show that the
commutation relations are satisfied it is necessary to use
\begin{equation}\label{3.14a}
\delta_{i}(\vec{x})T^{j}(\vec{y},
\gamma)=\delta_{i}^{j}\delta^{(2)}(\vec{x}-\vec{y}),
\end{equation}
which can be readily verified.

Using (\ref{3.14}) we can write down the covariant momenta as
\begin{eqnarray}\label{3.15}
\hat{{p}_{i}}-q\hat{A}_{i}(\vec{r})&\longrightarrow & -i
\left(\frac{\partial}{\partial
{r^{i}}}+q\delta_i(\vec{r})\right)\equiv -iD_{i}(\vec{r}),\\
\hat{{P}_{i}}-Q\hat{B}_{i}(\vec{R})&\longrightarrow &
-i\left(\frac{\partial}{\partial
{R^{i}}}+iQ\epsilon_{ij}T^{j}(\vec{R},\gamma)\right).
\end{eqnarray}
The gauge invariant combination $D_{i}(\vec{r})$ coincides with
the path derivative introduced by Mandelstam many years ago
\cite{MDS}. It comprises the ordinary derivative, representing the
momentum operator of the particle, plus $q$ times the ``path
derivative'' $\delta_{i}(\vec{r})$.
 With these realizations, the physical
constraints (\ref{3.10}) are written as
\begin{eqnarray}\label{3.16}
&&-\left({\rho}_1(x) +
\partial_i \Pi^{i}(x, \gamma)\right)\Psi[\gamma, \vec{r}, \vec{R}]\approx
0\quad \Longrightarrow \nonumber\\
&&\left(\delta^{(2)}(\vec{x}-\vec{r})-\sum_{s}(\delta^{(2)}(\vec{x}-\vec{b}_{s})-
\delta^{(2)}(\vec{x}-\vec{a}_{s}))\right)\Psi[\gamma, \vec{r},
\vec{R}]=0,
\end{eqnarray}
and
\begin{eqnarray}\label{3.17}
&&-\left({\rho}_2(x)+\epsilon^{ij}\partial_i
A_{j}(x)\right)\Psi[\gamma,\vec{r},\vec{R}]\approx 0
\quad \Longrightarrow \nonumber\\
&&\left({\rho}_2(x)
+\frac{i}{2}\epsilon^{ij}\Delta_{ij}(\vec{x})\right)\Psi[\gamma,\vec{r},\vec{R}]=0.
\end{eqnarray}
To write (\ref{3.16}) we have used
\begin{equation}\label{3.18}
\partial_i T^{i}(x,\gamma)\equiv -\varrho(\vec{x},\gamma)=-\sum_{s}(\delta^{(2)}(\vec{x}-\vec{b}_{s})-
\delta^{(2)}(\vec{x}-\vec{a}_{s})),
\end{equation}
with $\vec{a}_{s}$ and $\vec{b}_{s}$ labelling the starting and
ending points of the $s$-th ``strand'' of the path, respectively.

To solve the constraint (\ref{3.16}) we can use the geometrical
setting that allows to solve the Gauss constraint in Maxwell
theory coupled to non-relativistic particles \cite{E}. Following
this case, we consider ``Faraday`s lines'' states, consisting on
functionals that depend on an  open path composed  of $n$ strands
starting (or ending) at the particle`s position $\vec{r}$ . These
strands end (or start) at spatial infinity. To take into account
the source-free sector, this open strands might be accompanied by
closed contours too. For example dropping the contribution arising
from the starting points of the strands, the Gauss law constraint
(\ref{3.16}) can be written as,
\begin{eqnarray}\label{quant}
&&\left(q\delta^{(2)}(\vec{x}-\vec{r})-\sum_{s}
\delta^{(2)}(\vec{x}-\vec{a}_{s})\right)\Psi[\gamma,
\vec{r}, \vec{R}]=0\rightarrow\nonumber\\
&&\left(q\delta^{(2)}(\vec{x}-\vec{r})- n\,
\delta^{(2)}(\vec{x}-\vec{r})\right) \Psi[\gamma,\vec{r},
\vec{R}]=0.
\end{eqnarray}
This equation becomes an identity if $q=n$ for these incoming
paths (analogously $q=-n$ for outgoing ones). Is it easy to see
that for $N$ charges, one must take $N$ ``bundles'' of open paths,
one for each charged particle, having as many oriented strands so
the sum of incoming minus outgoing strands give the value of the
charge. Within this formalism there is no room for fractionary
charges, because a Faraday`s line carries a unit of electric flux,
which must be emitted from or absorbed by an integral charge $q$.
We find it convenient to denote the path-dependent functionals
that satisfy the Gauss constraint as $\Psi[\gamma_{\vec{r}},
\vec{R}]$, since this notation displays both the path and
point-dependence and recalls that from now on particles of "type
one" are attached to paths.

It should be observed that  Gauge invariant operators respect the
geometrical properties of the Faraday´s lines construction. For
instance, the "covariant momentum" $-iD_{i}(\vec{r})$ measures the
change of the wave-functional when both the particle and its
attached ``bundle'' of paths are infinitesimally displaced
\cite{E}.

Once the Gauss constraint (\ref{3.16}) is solved, we focus
ourselves in the second one (\ref{3.17}). Since
$\textbf{B}_{\vec{A}}=-\frac12\epsilon^{ij}F_{ij}\rightarrow
-\frac{i}{2}\epsilon^{ij}\Delta_{ij}(\vec{x})$, this constraint
tells us that each particle of "type two", whose position is
$\vec{R}$, carries an amount of ``magnetic flux'' proportional to
its electric charge, and confined to the point where the particle
``lives'' \cite{E}. We recognize this first class constraint as
the one that appears in the Maxwell-Chern-Simons theory when it is
quantized in the path representation \cite{LO}. Also, this
constraint appears in the path formulation of the theory of
particles interacting through a Chern-Simons field \cite{E}.

Following references \cite{LO} and \cite{E}, we can then write the
solution of (\ref{3.17}) as
\begin{equation}\label{mal}
\Psi[\gamma_{\vec{r}},
\vec{R}]=\exp\left(-i\frac{Q}{2\pi}\Theta(\gamma_{\vec{r}},\vec{R})
\right)\Phi(\partial\gamma_{\vec{r}},\vec{R}),
\end{equation}
with
$\epsilon^{ij}\Delta_{ij}(\vec{x})\Phi(\partial\gamma_{\vec{r}},
\vec{R})=0$ and
\begin{equation}
\frac12\epsilon^{ij}\Delta_{ij}(\vec{x})\Theta(\gamma_{\vec{r}},\vec{R})=\rho_2(x)
\end{equation}
The condition on $\Phi(\partial\gamma_{\vec{r}}, \vec{R})$ forces
it to be a function that depends on the path $\gamma_{\vec{r}}$
only through its boundary $\partial\gamma_{\vec{r}}=\vec{r}$. The
solution for $\Theta(\gamma_{\vec{r}},\vec{R})$ is the algebraic
sum of the angles subtended by the pieces (the strands) of the
path $\gamma$, measured from the point, $\vec{R}$, where the "big"
particle is
\begin{equation}\label{3.21}
\Theta(\gamma_{\vec{r}},\vec{R})\equiv\int_{\gamma}dx^{j}\,\varepsilon^{ij}
\left[\frac{(x-R)^{i}}{|\vec{x}-\vec{R}|^{2}}\right].
\end{equation}
It is interesting to remark a mayor difference between this case
and both the particles-Chern-Simons \cite{E} and
Maxwell-Chern-Simons \cite{LO} cases, concerning the constraint
(\ref{3.17}).  The present case does not suffers from what could
be called the "self-angle" problem. By this we refer to the fact
that in expression (\ref{3.21}), the angle subtended by the paths
is measured with respect to points that do not coincide with the
ending points of the paths. This contrasts with the Maxwell-CS and
particles-CS cases, where there appear self-interaction effects
that in the path representation manifest through the dependence of
the wave functional on the angle subtended by the path  measured
with respect to its own endpoints. This "self-angle" is ill
defined, and requires some regularizing prescription to deal with
it \cite{JC}.

 At this point one should verify whether the gauge
invariant operators of the theory preserve the form of the
physical states $\Psi[\gamma_{\vec{r}},
\vec{R}]=\exp\left(-i\frac{Q}{2\pi}\Theta(\gamma_{\vec{r}},\vec{R})\right)
\Phi({\vec{r}},\vec{R})$. For instance, let us consider the action
of the Mandelstam derivative onto the gauge invariant states. It
is given by
\begin{eqnarray}\label{3.23}
-iD_{i}(\vec{r})\Psi_{Phys}&=&-iD_{i}(\vec{r})\left[\exp\left(-i\frac{Q}{2\pi}
\Theta(\gamma_{\vec{r}},\vec{R})\right)\Phi({\vec{r}},
\vec{R})\right]\nonumber\\
&=&\exp\left(-i\frac{Q}{2\pi}
\Theta(\gamma_{\vec{r}},\vec{R})\right)\left[-i\frac{\partial}{\partial
{r^{i}}}+\frac{qQ}{2\pi}\epsilon_{ij}\frac{(r-R)^{j}}{|\vec{r}-\vec{R}|^2}\right]\Phi({\vec{r}},
\vec{R})\nonumber\\&=&\exp\Big(-i\frac{Q}{2\pi}
\Theta(\gamma_{\vec{r}},\vec{R})\Big)\Phi'({\vec{r}},\vec{R}),
\end{eqnarray}
where the second line defines $\Phi'(\vec{r},\vec{R})$. Hence, we
see that the Mandelstam derivative leaves invariant the physical
space of states, as it should be.

On the other hand, it can be verified that the other ``covariant
momentum''
\begin{equation}\label{3.24}
P_i-QB_i(\vec{R})\rightarrow -i\left(\frac{\partial}{\partial
{R^{i}}}+iQ\epsilon_{ij}T^{j}(\vec{R},\gamma)\right),
\end{equation}
produces a result analogous to (\ref{3.23}) when applied to the
physical functionals
\begin{eqnarray}
-i\left(\frac{\partial}{\partial
{R^{i}}}+iQ\epsilon_{ij}T^{j}(\vec{R},\gamma)\right)
\Psi_{Phys}&=&\nonumber \\
&=&\exp\left(-i\frac{Q}{2\pi}
\Theta(\gamma_{\vec{r}},\vec{R})\right)\left[-i\frac{\partial}{\partial
{R^{i}}}-\frac{qQ}{2\pi}\epsilon_{ij}\frac{(r-R)^{j}}{|\vec{r}-\vec{R}|^2}\right]\Phi({\vec{r}},
\vec{R})\nonumber\\&=&\exp\Big(-i\frac{Q}{2\pi}
\Theta(\gamma_{\vec{r}},\vec{R})\Big)\Phi''({\vec{r}},\vec{R}),
\end{eqnarray}
where we have taken into account the quantization condition
(\ref{quant}) for q. Again, gauge invariance is maintained.

As in the Maxwell-Chern-Simons and particles-Chern-Simons cases
\cite{LO,E} there is a unitary transformation that allows us to
eliminate the path dependent phase factor
$\exp\Big(-i\frac{Q}{2\pi}\Theta(\gamma_{\vec{r}},\vec{R})\Big)$.
Once this transformation is performed, the path dependence of the
wave functional is reduced to the boundary of the path
$\partial\gamma_{\vec{r}}$, which is just the set of the positions
$\{\vec{r}\}$ of the type-one particles. At this point, the
boundary dependence of the wave functional becomes redundant, and
it suffices to employ ordinary (i.e., point-dependent) wave
functions $\Psi(\vec{r},\vec{R})$. At the same time it is not
necessary to maintain the path (or loop) derivatives in the
physical operators, and we may substitute them by  ordinary
derivatives\cite{E}.

Once this unitary transformation is performed, the Schr\"{o}dinger
equation of the model can be written down as
\begin{equation}\label{3.25}
i\partial_{t}\Psi(\vec{r},t)=H_{0}\Psi(\vec{r},t),
\end{equation}
where the Hamiltonian $H_0$ is
\begin{equation}\label{3.26}
H_0=\frac{mv^2}{2}+ \frac{MV^{\,2}}{2},
\end{equation}
where $mv^i$ and $MV^{i}$ act on the states as
\begin{eqnarray}\label{3.27}
m{v^{i}}=-i\frac{\partial}{\partial {r^{i}}}-eqA_i(\vec{r})
={p_{i}}+\frac{qQ}{2\pi}\epsilon_{ij}\frac{(r^{j}-R^{j})}{|\vec{r}-\vec{R}|^2} \\
M{V^{i}}=-i\frac{\partial}{\partial
{R^{i}}}-QB_i(\vec{R})={P_{i}}-\frac{qQ}{2\pi}\epsilon_{ij}
\frac{(r^{j}-R^{j})}{|\vec{r}-\vec{R}|^2}.
\end{eqnarray}
Thus we arrive to the quantum mechanics of two species of
non-relativistic particles that interact through potentials that
satisfy
\begin{equation}\label{3.27a}
qA_{i}(\vec{r})=-QB_{i}(\vec{R})=-\frac{qQ}{2\pi}\epsilon_{ij}
\frac{(r^{j}-R^{j})}{|\vec{r}-\vec{R}|^2}
\end{equation}
It is straightforward to see that this potentials solve the
constrains $\chi_{1}(x)$ and $\chi_{2}(x)$ as in (\ref{3.7}). This
long-range interaction coincides with the topological interaction
experienced by anyons \cite{Wilczek,Wu,Jackiw,Indio,Arovas}. In
fact, we have recovered the hamiltonian of precisely two anyons,
which is exactly soluble \cite{Indio,Leinas,Wu,Wilczek}.

It is worth recalling that the equations describing anyons  can be
rewritten in what some authors call the "anyon gauge". It is
obtained by performing a singular gauge transformation that
converts the Schrodinger equation for the topologically
interacting particles into that of a free-particles system.
However, the interaction remains hidden in the fact that the wave
function becomes multivalued. A moment of reflection allows to see
that this features are neatly realized in the path-dependent
formulation: the wave function in the anyon gauge
\begin{equation}\label{3.27b}
\Psi(\vec{r},t)=\exp\Big(-i\frac{Q}{2\pi}
\Theta(\gamma_{\vec{r}},\vec{R})\Big)\Phi(\vec{r},t),
\end{equation}
precisely corresponds to the multivalued wave function
(\ref{mal}), while the Mandelstam derivative is just the "partial
derivative" appropriate to act onto that multivalued  wave
function, that turned to be a path-dependent one.

\section{$3+1$ dimensional BF theory and non-relativistic string-particle interaction}\label{sec4}

Our last model consists on a dynamical string interacting with a
dynamical particle by means of a topological BF term in $3+1$
dimensions. Both string and particle are non-relativistic, and are
described as in the preceding sections. For the sake of simplicity
we restrict ourselves to consider just one particle and one
string, although the formulation could certainly be extended to a
more general case. The BF term is analogous to its counterpart in
$2+1$ dimensions studied in the last section. We take the action
as
\begin{eqnarray}\label{4.1}
S&=&\frac14\int
d^{4}x\epsilon^{\mu\nu\lambda\rho}B_{\mu\nu}F_{\lambda\rho}+ \int
dt \,\left(\frac12 \,m\dot{\vec{r}}^{\,2}\right)+
\frac{\alpha}{2}\int dt \int d\sigma \left[(\dot{z}^i)^2-(z'^i)^2)
\right]\nonumber\\&+&\int d^4x \left(\textit{J}^{\mu}A_{\mu}
+\frac{1}{2}\textit{J}^{\mu\nu}B_{\mu\nu}\right).
\end{eqnarray}
As before, we have
\begin{eqnarray}\label{4.2}
\textit{J}^{\mu}(\vec{x},t)&=&q\int
dy^{\mu}\delta^{(4)}(x-y)=qv^{\mu}(t)\delta^{(3)}(\vec{x}-\vec{r})
\equiv(\rho(x),\vec{J}(x)),\\ \textit{J}^{\mu\nu}(\vec{x},t)&=&
\phi \int dt \int d\sigma \left[\dot{z}^\mu z'^\nu -\dot{z}^\nu
z'^\mu\right]\delta^{(4)}(x-z),
\end{eqnarray}
where $v^{\mu}(t)=(1,\vec{v})$, and
$F_{\mu\nu}=\partial_{\mu}A_{\nu}-\partial_{\nu}A_{\mu}$. As in
the preceeding sections, $\alpha$ is the string tension, and
$\phi$ and $q$ are dimensionless, so the units of the fields are
clear in this context. The $3+1$ decomposition of the action
yields
\begin{eqnarray}\label{4.3}
&S&=\frac{1}{2}\int d^4x
\left[\dot{A}_{i}\epsilon^{ijk}B_{jk}+B_{0i}\epsilon^{ijk}F_{jk}+A_{0}\epsilon^{ijk}\partial_{i}B_{jk}\right]\nonumber\\
&+&\int
dt\,\left(\frac12\,m\dot{\vec{r}}^{\,2}\right)+\frac{\alpha}{2}\int
dt \int d\sigma\left[(\dot{z}^i)^2-(z'^i)^2)
\right]\nonumber\\&+&q\int dt\left[ \dot{r}^i(t)A_i(\vec{r}(t)) +
A_{0}(\vec{r}(t))\right]+\phi\int dt \int
d\sigma\left[B_{ij}(z(\sigma,t))\dot{z}^{i}z'^{j}+
B_{0k}(z(\sigma,t))z'^{k}\right]\nonumber,
\end{eqnarray}
[$z^{0}=t$, $\dot{z}^{0}=1$, $z'^{0}=0$]. The expressions
${A}_{\mu}(\vec{r}(t),t)\equiv{A}_{\mu}(\vec{r})$ and
$B_{\mu\nu}(z(\sigma,t))\equiv B_{\mu\nu}(z)$ should be understood
as a shorthand for
\begin{eqnarray}\label{4.4}
A_{\mu}(\vec{r},t)\equiv \int
d^{3}\vec{x}\;\delta^{3}(\vec{x}-\vec{r})A_{\mu}(\vec{x},t),\\
B_{\mu\nu}(z(\sigma,t))\equiv\int
d^{3}\vec{x}\;\delta^{3}(\vec{x}-\vec{z})B_{\mu\nu}(\vec{x},t).
\end{eqnarray}

We now summarize the Dirac quantization of the theory. The
conjugate momenta associated to the  particle and string are given
by
\begin{eqnarray}\label{4.5}
&&{\Pi^{k}}_{(A)}=\frac{1}{2}\epsilon^{ijk}B_{ij},\\
&&p_{i}=m\frac{dr^{i}}{dt}+qA_{i}(\vec{r})\longrightarrow
\dot{r}^{i}=\frac{(p_i-q A_i(\vec{r}))}{m},\\
&&P_{i(z)}=\alpha \dot{z}^{i}+\phi B_{ij}(z){z'}^j\longrightarrow
\dot{z}^{i}=\frac{(P_{i(z)}-\phi B_{ij}(z){z'}^j)}{\alpha}.
\end{eqnarray}
On the other hand, and as in the preceding section, it should be
noticed that $A_{i}$ and $\frac{1}{2}\epsilon^{ijk}B_{jk}$ are
already canonical conjugate variables. Also, we will not take
$B_{i0}$ and $A_0$ as dynamical variables from the beginning.

 The Hamiltonian
can be written as follows
\begin{eqnarray}\label{4.6}
H&=&\frac{(p_i-q A_i(\vec{r}))^2}{2m}+\int d\sigma
\frac{\alpha}{2}\left[\frac{1}{\alpha^2}\left(P_{i(z)}-\phi
B_{ij}(z)z'^{j}\right)^{2}+(z'^{i})^2 \right] +\int
d^3x\,A_{0}(x)\chi (x)\nonumber\\
&&+\int d^3x\,B_{0j}(x)\chi^j (x)\nonumber \\
&\equiv& H_0 + \int d^3x\,A_{0}(x)\chi (x)+\int
d^3x\,B_{0j}(x)\chi^j (x).
\end{eqnarray}
So $B_{i0}$ and $A_0$ appear as the Lagrange multipliers
associated to the first class constraints that generate time
independent gauge transformations
\begin{eqnarray}\label{4.7}
&&\chi (x)\equiv -\rho(x)-\frac{1}{2}\epsilon^{ijk}\partial_{i}B_{jk} \\
&&\chi^i (x)\equiv - \rho^{i}(x) - \epsilon^{ijk}\partial_j
A_{k}(x)),
\end{eqnarray}
with
\begin{eqnarray}\label{4.8}
&&\rho(x)=q\delta^{(3)}(\vec{x}-\vec{r}),\\
&&\rho^{i}(x)\equiv \phi \int d\sigma
z'^i\delta^{(3)}(\vec{x}-\vec{z}).
\end{eqnarray}
The canonical Poisson brackets are defined as
\begin{equation}\label{4.9}
\left\{r^i,P_j \right\}=\delta^i_j,
\end{equation}
\begin{equation}\label{4.10}
\left\{z^i(\sigma),P_j(\sigma')\right\}=\delta^i_j\delta(\sigma
-\sigma'),
\end{equation}
\begin{equation}\label{4.11}
\left\{A_{i}(\vec{x}),\frac{1}{2}\epsilon^{jkl}B_{kl}(\vec{y})\right\}=\delta_{i}^{j}
\delta^{(3)}(\vec{x}-\vec{y}).
\end{equation}
The remaining Poisson brackets vanish.  Due to the topological
character of the BF term the contribution of the fields to the
Hamiltonian looks as if they were external fields, just like in
the preceding section. The basic observables, in Dirac`s sense,
are the positions of the particle and string, $\vec{r}$ and
 $\vec{z}(\sigma)$, and the ``covariant'' (gauge invariant) momenta
of the particle and the string, given by
\begin{eqnarray}\label{4.12}
p_i- q A_i(\vec{r}),\qquad P_{i(z)}-\phi B_{ij}(z)z'^{j},
\end{eqnarray}
respectively. All the physical observables of the theory are built
in terms of these gauge invariant quantities. For instance, the
Hamiltonian fulfils this rule.

To quantize, the canonical variables are promoted to operators
obeying the commutators that result from the usual replacement
$\{\;,\;\}\;\longrightarrow\;-i[\;,\;]$. These operators have to
be realized in a Hilbert space of physical states
$|\Psi\rangle_{Phys}$ that obey
\begin{eqnarray}\label{4.14}
-\left(\rho(x)+\frac{1}{2}\epsilon^{ijk}\partial_{i}B_{jk}\right)
|\Psi\rangle_{Phys}\approx 0
\end{eqnarray}
\begin{eqnarray}\label{4.14a}
-\left(\rho^{i}(x) + \epsilon^{ijk}\partial_j
A_{k}(x))\right)|\Psi\rangle_{Phys}\approx 0
\end{eqnarray}
Now we seek for a geometric representation appropriate to the
present model. As we shall discuss, there are two possible
choices, depending on which of the field operators ($A_{\mu}(x)$
or $B_{\mu\nu}(x)$) we take as "position" or as "derivative"
operator. In both choices, we shall realize  the operators
associated to the particle and string in a Schrodinger or ``shape"
representation, i.e., we shall take
\begin{eqnarray}\label{4.15}
\hat{r}^i \longrightarrow r^i, \qquad \hat{z}^i(\sigma)
\longrightarrow z^i(\sigma),\\\hat{P}^i \longrightarrow
-i\frac{\partial} {\partial r^i},\qquad
\hat{P}^i(\sigma)\longrightarrow -i \frac{\delta}{\delta
z^i(\sigma)}.
\end{eqnarray}
These operators are supposed to act onto functionals $\Psi
[\vec{r}, z(\sigma)]$ that depend on the coordinates of the
particle and of the string world-sheet. Once the particle and
string operators are realized, we have also to accommodate the
fields operators into the description. The first geometric
representation that we are going to consider is a ``Faraday`s
lines'' or path representation \cite{GT}. We begin observing that
the fundamental commutator associated to equation (\ref{4.11}) can
be realized on path-dependent functionals if one prescribes
\begin{eqnarray}\label{4.16}
&&\hat{A}_{i}(\vec{x}) \longrightarrow i\delta_{i}(\vec{x}),\nonumber\\
&&\hat{\Pi}^{i}=\frac{1}{2}\epsilon^{ijk}\hat{B}_{jk}(\vec{x})\longrightarrow
T^{i}(\vec{x},\gamma),
\end{eqnarray}
where $\delta_{i}(\vec{x})$ and $T^{i}(\vec{x},\gamma)$ were
defined in (\ref{3.11}) and (\ref{3.12}). In this representation
we can write
\begin{eqnarray}\label{4.17}
\hat{p}_i- q \hat{A}_i(\vec{r})\longrightarrow
-i\Big(\frac{\partial} {\partial
r^i}+q\delta_{i}(\vec{r})\Big)\equiv
-iD_{i}(\vec{r}),\nonumber\\
\hat{P}_{i(z)}-\phi \hat{B}_{ij}(z)\hat{z'}^{j}\longrightarrow
-i\Big(\frac{\delta}{\delta z^i(\sigma)}-i\phi\epsilon_{ijk} z'^j
T^{k}(z(\sigma),\gamma)\Big),
\end{eqnarray}
where $D_{i}(\vec{x})$ is the ``Mandelstam operator'' as defined
in (\ref{3.15}). Thus, the states of the interacting theory can be
taken as functionals $\Psi[\vec{r},z(\sigma),\gamma]$. Among them,
the physical ones will be those that satisfy the constraints
(\ref{4.14}) and (\ref{4.14a}). The former can be written down as
\begin{eqnarray}\label{4.18}
&&-\left(\rho(x)+\frac{1}{2}\epsilon^{ijk}\partial_{i}B_{jk}\right)
\Psi[\vec{r},z(\sigma),\gamma]\approx 0 \qquad\Longrightarrow \nonumber\\
&&\left(\rho(x)+\partial_{i}T^{i}(\vec{x},\gamma)\right)\Psi[\vec{r},z(\sigma),\gamma]=\left(
q\delta^{(3)}(\vec{x}-\vec{r})-\varrho
(\vec{x},\gamma)\right)\Psi[\vec{r},z(\sigma),\gamma]= 0,
\end{eqnarray}
where $\varrho(\vec{x},\gamma)$ was defined before in
(\ref{3.18}).

Regarding the second constraint, we have
\begin{eqnarray}\label{4.19}
&&-\left(\rho^{i}(x) + \epsilon^{ijk}\partial_j
A_{k}(x)\right)\Psi[\vec{r},z(\sigma),\gamma]\approx
0\Longrightarrow \nonumber\\&& \left(\rho^{i}(x)+i
\frac{1}{2}\epsilon^{ijk}\Delta_{jk}(\vec{x})\right)
\Psi[\vec{r},z(\sigma),\gamma]= 0,
\end{eqnarray}
with $\Delta_{ij}(\vec{x})$ given in section \ref{sec3}. At this
point it will be useful to recall the solution of the constraints
in the "toy model". We see that (\ref{4.18}) is similar to
(\ref{3.16}), while (\ref{4.19}) corresponds to a generalized
version of (\ref{3.17}). From our experience with those
constraints, we obtain the following picture: (\ref{4.18}) tells
us that we have to take as physical wave functions those that
depend on an open-path of  $n$-strands that meet at  the point
$\vec{r}$ where the charged particle is located (as before, this
open path may also comprise closed pieces). The number of oriented
strands " sum up" to the (quantized) value of the electric charge.
These "Faraday´s lines" drawing has to be accompanied by a closed
string, which has nothing attached in this representation. From
now on we shall write $\gamma_{\vec{r}}$ instead of $\gamma$,
because gauge invariance joins paths and particles as explained
before. On the other hand, following the solution of (\ref{3.17})
we see that (\ref{4.19}) obligates to write the physical wave
functionals as
\begin{equation}\label{4.20}
\Psi[\vec{r},\vec{z}(\sigma),\gamma_{\vec{r}}]=\exp\left(i\,\Theta(\vec{r},
z(\sigma),\gamma_{\vec{r}})\right)\Phi(\vec{r},\vec{z}(\sigma)),
\end{equation}
with
\begin{equation}\label{coleada}
\frac12\epsilon^{ijk}\Delta_{jk}(\vec{x})\Theta(\vec{r},
z(\sigma),\gamma_{\vec{r}})= \rho^{i}(\vec{x}).
\end{equation}
The solution of (\ref{coleada}) is
\begin{equation}\label{4.21}
\Theta(\vec{r},
z(\sigma),\gamma_{\vec{r}})=\frac{\phi}{4\pi}\int_{\gamma} dx^i
\int_{\Gamma} d\sigma z'^{j}
\epsilon_{ijk}\frac{(x-z(\sigma))^k}{|\vec{x}-\vec{z}(\sigma)|^3},
\end{equation}
where $\gamma$ is the path (as usual), and $\Gamma$ is the closed
curve that coincides with the closed string.

Thus, we obtain that the physical wave function comprises a fixed
path-dependent phase factor times $\Phi$, which depends on the
path $\gamma$ only through its ending point (the one which is not
at spatial infinity), that is where the particle ``lives''.
 In order to give a physical interpretation of
$\Theta(\gamma)$, we use Stokes Theorem and the fact that
\begin{equation}\label{4.22}
\frac{(x-z(\sigma))^k}{|\vec{x}-\vec{z}(\sigma)|}=
\frac{\partial}{\partial
z^k}\left[\frac{1}{|\vec{x}-\vec{z}(\sigma)|}\right]=-\frac{\partial}{\partial
x^k}\left[\frac{1}{|\vec{x}-\vec{z}(\sigma)|}\right]
\end{equation}
to rewrite (\ref{4.21}) in the  form
\begin{equation}\label{4.23}
\Theta(\vec{r},
z(\sigma),\gamma_{\vec{r}})=\phi\left[\frac{1}{4\pi}
\int_{\Sigma(\Gamma)}dS_{i}\,\left[\frac{(b-z)^i}{|\vec{b}-\vec{z}|^3}
-\frac{(a-z)^i}{|\vec{a}-\vec{z}|^3}
\right]-\int_{\Sigma(\Gamma)}dS_{i}
\int_{\gamma}dx^i\delta^{(3)}(\vec{x}-\vec{z})\right],
\end{equation}
with $dS_{i}\equiv\epsilon_{ijk}d\Sigma^{jk}$  ($d\Sigma^{jk}$ was
defined in eq. (\ref{2.16})). In the last expression
$\Sigma(\Gamma)$ is an open surface that has the string $\Gamma$
as its border, i.e., $\partial\Sigma{(\Gamma)}=\Gamma$. The first
term in (\ref{4.23}) is the solid angle subtended by the surface
$\Sigma{(\Gamma)}$ attached to the string measured from the final
point  $\vec{b}$, minus the solid angle subtended by the same
surface but measured from the starting point $\vec{a}$ of the path
$\gamma$ [again, it should be recalled that one of these points is
at spatial infinity]. In turn, the second term in (\ref{4.23})
counts the number of times that the path $\gamma$ intersects the
surface $\Sigma{(\Gamma)}$. Although $\Theta(\vec{r},
z(\sigma),\gamma_{\vec{r}})$, given in equation (\ref{4.23}) looks
like  a surface-dependent quantity, this dependence is only
apparent, as can be realized by just turning back to (\ref{4.21}).

It is worth comparing this case with the toy model of the
preceding section. In the toy model the geometrical phase
analogous to $\Theta(\vec{r}, z(\sigma),\gamma_{\vec{r}})$
measured the winding number of the path attached to one of the
particles, around the other particle in the plane. In the present
case, $\Theta(\vec{r}, z(\sigma),\gamma_{\vec{r}})$ generalizes
this geometrical fact to a three dimensional situation: it
measures the "winding" of the path attached to the particle
"around" the closed string. There is yet another possibility of
generalizing this in three dimensions. It corresponds precisely to
the other geometric representation, that we next discuss.

 The second, ``dual''
representation is, in fact, a surface-dependent representation, as
the one discussed in section \ref{sec2}. We set
\begin{eqnarray}\label{4.24}
\hat{A}_{i}(\vec{x}) \longrightarrow &&-\,\frac12\epsilon_{ijk}
T^{jk}(\vec{x},\Sigma),\nonumber\\
\hat{\Pi}^{i}(\vec{x}) \longrightarrow&&
i\epsilon^{ijk}\delta_{jk}(\vec{x})\quad\Longrightarrow\\
\hat{B}_{ij}(\vec{x})\longrightarrow &&
2i\delta_{ij}(\vec{x}),\nonumber
\end{eqnarray}
where $T^{jk}(\vec{x},\Sigma)$ and $\delta_{ij}(\vec{x})$ were
defined in (\ref{2.16}) and (\ref{2.17}).  It can be readily seen
that this prescriptions realize the canonical commutators of the
theory. Also, a straightforward calculation from (\ref{4.7}),
using (\ref{4.24}) and (\ref{2.19}) leads to write the first class
constraints as,
\begin{eqnarray}\label{4.25}
\chi(\vec{x})=-\rho(\vec{x})-\frac{i}{3}\epsilon^{ijk}\triangle_{ijk}=0,
\end{eqnarray}
and
\begin{eqnarray}\label{4.26}
\chi^{i}(\vec{x})&=&-\rho^{i}(\vec{x})+\epsilon^{ijk}\partial_{j}\Big(\frac12
\epsilon_{klm}T^{lm}(\vec{x},\Sigma)\Big)=-\rho^{i}(\vec{x})+T^{i}(\vec{x},\partial\Sigma)=0\quad
\longrightarrow \nonumber\\&&-\phi\int_{\Gamma}
dz^{k}\delta^{(3)}(\vec{x}-\vec{z}_{(\sigma)})
+\int_{\gamma=\partial\Sigma}
dy^{k}\delta^{(3)}(\vec{x}-\vec{y})=0.
\end{eqnarray}
Equation (\ref{4.26}) tells us that the string should coincide
with the boundary of the open surface (see \cite{E} and the
discussions of section \ref{sec2}). On the other hand, our
previous experience with the toy model and with the former
representation for the present model also teaches us that the
solution of (\ref{4.25}) is given by wave functionals of the form
(see \cite{LO,E} and the discussion of section \ref{sec3})
\begin{equation}\label{4.27}
\Psi[\vec{r},z,\Sigma]=\exp\Big(i\,\Theta(\vec{r},
\vec{z},\Sigma)\Big)\Phi(\vec{r},\vec{z}),
\end{equation}
where
\begin{equation}\label{4.28}
\Theta(\vec{r},
\vec{z},\Sigma)=\frac{q}{8\pi}\int_{\Sigma(\Gamma)}dS_{i_y}
\frac{(y-r)^i}{|\vec{y}-\vec{r}|^3},
\end{equation}
is proportional to the solid angle subtended by the surface $\Sigma(\Gamma)$
measured from the position of the particle $\vec{r}$. In analogy
with the toy model and with the path representation of this model,
we see that the dependence on the surface is restricted to a phase
factor, which measures a topological feature: how many times the
surface attached to the string wraps around the particle.

So, in the surface representation we end up with strings having a
bundle of $n$ pieces of open-surfaces attached to them, with $n$
depending of the value of the quantized constant $\phi$ ( i.e.,
the ``charge'' of the string). Also, the wave functional depends
of a``lonely" point charged particle. The role of the surface is
to take into account how the the string and particle are
topologically related. It could be said that the difference
between the two dual representations is encoded in the following
feature: which of the matter objects (the particle or the string)
is left alone, and which has an attached object whose winding or
wrapping around the other carries the content of the topological
interaction.

It is interesting to see how  gauge invariance is maintained
through a geometric mechanism, in both representations . For
instance, in the path representation (\ref{4.16}), the
``covariant'' momentum associated with the particle (\ref{4.12})
is again realized as a ``Mandelstam'' operator that translates
both the particle and its attached  "bundle of paths" together
(see discussion in section \ref{sec3}). Also, in the surface
representation (\ref{4.24}) where the string is coupled to the
"bundle of surfaces", the covariant momentum of the string
(expression (\ref{4.12})) translates both the string and the set
of surfaces together, thus maintaining the geometrical picture
dictated by gauge invariance (see also the discussion at the end
of the first section)
\begin{equation}\label{4.29}
P_{i(z)}-\phi
B_{ij}(z)z'^{j}\;\longrightarrow\;-i\left(\frac{\delta}{\delta
z^{i}}+2\phi\delta_{ij}(\vec{z})z'^{j}\right).
\end{equation}
The last expression is a kind of generalized ``Mandelstam
operator'' for the string-surface representation.

On the other hand, gauge invariance also restricts the form of the
path (or surface) dependent wave functional, accordingly with
(\ref{4.20}) or (\ref{4.27}). We should check that the observables
of the theory respect this particular form. To this end, we apply
the gauge invariant momenta to the physical states $\Psi_{Pys}$.
In the path representation we obtain
\begin{eqnarray}\label{4.30}
p_i-q A_i(\vec{r})\;\rightarrow &&-i\left(\frac{\partial}{\partial
r^{i}}+q\delta_{i}(\vec{r})\right)\Psi_{Pys}(\vec{r},\vec{z},\gamma_{\vec{r}})=
\nonumber\\\qquad&&=\exp(i\,\Theta)\left[-i\frac{\partial}{\partial
r^{i}}+\frac{q\phi}{4\pi}\int_{\Gamma}dz^{j}\epsilon_{ijk}
\frac{(r-z)^k}{|\vec{r}-\vec{z}|^3}\right]\Phi(\vec{r},\vec{z})\nonumber\\
\qquad&&=\exp\left[i\,\Theta(\vec{r},
z(\sigma),\gamma_{\vec{r}})\right]\times\Phi'(\vec{r},\vec{z}),
\end{eqnarray}
and using the constraint (\ref{4.18}) we get
\begin{eqnarray}\label{4.31}
P_{i(z)}-\phi B_{ij}(z)z'^{j}\;\rightarrow &&
-i\left(\frac{\delta}{\delta
z^{i}}-i\phi\epsilon_{ijk}z'^{j}T^{k}(\vec{z},\gamma)\right)
\Psi_{Pys}(\vec{r},\vec{z},\gamma_{\vec{r}})=\nonumber\\
\qquad&&=\exp(i\,\Theta)\left[-i\frac{\delta}{\delta
z^{i}}+\frac{q\phi}{4\pi}\epsilon_{ijk}z'^{j}
\frac{(r-z)^k}{|\vec{r}-\vec{z}|^3}\right]\Phi(\vec{r},\vec{z})\nonumber\\
\qquad&&=\exp\left[i\,\Theta(\vec{r},
z(\sigma),\gamma_{\vec{r}})\right]\times\Phi'(\vec{r},\vec{z}).
\end{eqnarray}

In turn, in the surface representation we have
\begin{eqnarray}\label{4.32}
p_i-q A_i(\vec{r})\;\rightarrow &&-i\left(\frac{\partial}{\partial
r^{i}}+i\frac{q}{2}\epsilon_{ijk}T^{jk}(\vec{r},\Sigma)\right)\Psi_{Pys}(\vec{r},\vec{z},\Sigma)=\nonumber\\
\qquad&&=\exp(i\,\Theta)\left[-i\frac{\partial}{\partial
r^{i}}+\frac{q\phi}{4\pi}\int_{\Gamma}dz^{j}\epsilon_{ijk}
\frac{(r-z)^k}{|\vec{r}-\vec{z}|^3}\right]\Phi(\vec{r},\vec{z})\nonumber\\
\qquad&&=\exp\left[i\,\Theta(\vec{r},
z(\sigma),\Sigma)\right]\times\Phi'(\vec{r},\vec{z}),
\end{eqnarray}
where we have used the quantization constraint (\ref{4.26}). For
the other gauge invariant operator we have
\begin{eqnarray}\label{4.33}
P_{i(z)}-\phi B_{ij}(z)z'^{j}\;\rightarrow &&
-i\left(\frac{\delta}{\delta
z^{i}}+2\phi\delta_{ij}(\vec{z})z'^{j}\right)
\Psi_{Pys}(\vec{r},\vec{z},\Sigma)=\nonumber\\
\qquad&&=\exp(i\,\Theta)\left[-i\frac{\delta}{\delta
z^{i}}+\frac{q\phi}{4\pi}\epsilon_{ijk}z'^{j}
\frac{(r-z)^k}{|\vec{r}-\vec{z}|^3}\right]\Phi(\vec{r},\vec{z})\nonumber\\
\qquad&&=\exp\left[i\,\Theta(\vec{r},
z(\sigma),\Sigma)\right]\times\Phi'(\vec{r},\vec{z}).
\end{eqnarray}

 In all these expressions, the functionals
 " $\Phi'$ " only depend on the particle and string
 positions. Hence, the observables leave invariant the physical
 sector of the Hilbert space, as required. Furthermore, these expressions indicate that the path
or the surface dependence may be eliminated by performing a
unitary transformation that extracts from the wave functional the
phase factor  $\exp\left[i\,\Theta(\vec{r},
z(\sigma),\gamma_{\vec{r}})\right]$ or
$\exp\left[i\,\Theta(\vec{r}, z(\sigma),\Sigma)\right]$ . This
unitary transformation appeared in similar contexts in
\cite{E,LO}.

It is interesting to see how the hamiltonian looks before
performing the unitary transformation mentioned above. In the path
representation we have, introducing (\ref{4.17}) in (\ref{4.6})

\begin{eqnarray}\label{4.34}
H_0=\frac{\left[-i\left(\frac{\partial}{\partial
r^{i}}+q\delta_{i}(\vec{r})\right)\right]^2}{2m}+\int
d\sigma\frac{\alpha}{2}\left[\frac{\left(-i\frac{\delta}{\delta
z^{i}}-\phi\epsilon_{ijk}z'^{j}T^{k}(\vec{z},\gamma)\right)^2}{\alpha^2}+(z'^{i})^2\right].
\end{eqnarray}

On the other hand, by a similar calculation, in the surface
representation the hamiltonian would be given by

\begin{eqnarray}\label{4.34}
H_0=\frac{\left[-i\left(\frac{\partial}{\partial
r^{i}}+i\frac{q}{2}\epsilon_{ijk}T^{jk}(\vec{r},\Sigma)\right)\right]^2}{2m}+\int
d\sigma\frac{\alpha}{2}\left[\frac{\left(-i\frac{\delta}{\delta
z^{i}}-2i\phi\delta_{ij}(\vec{z})z'^{j}\right)^2}{\alpha^2}+(z'^{i})^2\right].
\end{eqnarray}

Both expresions yield the same Schrodinger equation

\begin{eqnarray}\label{4.35}
i\frac{\partial}{\partial t}\Psi [\vec{r},\vec{z}]&=&H_{0}\Psi
[\vec{r},\vec{z}]\nonumber\\
&=&\left\{\frac{1}{2m}\left[-i\frac{\partial}{\partial
r^{i}}+\frac{q\phi}{4\pi}\int_{\Gamma}dz^{j}\epsilon_{ijk}
\frac{(r-z)^{k}}{|\vec{r}-\vec{z}|^{3}}\right]^{2}\right.\nonumber\\&&\left.
+\int d\sigma
\frac{\alpha}{2}\left[\frac{\left(-i\frac{\delta}{\delta
z^{i}}+\frac{q\phi}{4\pi}\epsilon_{ijk}z'^{k}\frac{(r-z)^{j}}{|\vec{r}-\vec{z}|^{3}}\right)^{2}}{\alpha^{2}}+
(z'^{i})^{2}\right]\right\}\times \Phi(\vec{r},\vec{z}),
\end{eqnarray}
once the unitary transformation is performed. The last equation is
the analogous of the equation for the system of two anyons
(\ref{3.27}) that the toy model yielded. The right hand side
corresponds to the energy of a particle and a string that interact
through a topological generalized potential of the form

\begin{equation}\label{4.36}
A_{i}(\vec{r},\vec{z})=\frac{q\phi}{4\pi}\int_{\Gamma}d\sigma\epsilon_{ijk}z'^{j}
\frac{(r-z)^{k}}{|\vec{r}-\vec{z}|^{3}}.
\end{equation}

This suggests that there should be an equivalent formulation of
the model, that only deals with particle and string variables (and
not with topological fields) from the very beginning. In fact, it
is easy to see that the lagrangean

\begin{equation}\label{4.37}
L=\frac{1}{2}m\dot{\vec{r}}^{2}+\int
d\sigma\frac{\alpha}{2}\left[(\dot{z}^{i})^{2}-(z'^{i})^{2}\right]+
\frac{q\phi}{4\pi}\int_{\Gamma}d\sigma\epsilon_{ijk}z'^{j}
\frac{(r-z)^{k}}{|\vec{r}-\vec{z}|^{3}}(\dot{r}^{i}-\dot{z}^{i}),
\end{equation}
fulfils this requisite.

% Notice that the last term in (\ref{4.37})
%may be written as the usual $\int d\sigma
%\vec{A}\cdot(\dot{\vec{r}}-\dot{\vec{z}})$ term.

\section{Discussion}

We have studied the geometric representation of strings
interacting by means of the Kalb-Ramond field. We saw that this
representation is a ``surface representation'' that may be set up
only if the coupling constant $\phi$ of the string (equivalent to
the charge if they were point particles) is quantized as integer
values $n$. This theory is in a sense very similar to the Maxwell
theory interacting with $0$-dimensional objects studied in the
framework of the LR in \cite{E}. In this case the quantization
within the LR was a ``Faraday`s lines representation'' where the
quantization of the charges was stated in terms of the fundamental
unit of electric flux carried by each Faraday`s line. In both
cases the appropriate Hilbert space is made of wave functionals
whose arguments are geometric ``Faraday`s extended objects'' (that
in this work are surfaces) emanating from or ending at the strings
(or particles) positions.

We also studied two generalizations of the path-space formulation
of the  theory of particles interacting through a Chern-Simons
field  \cite{E}. First, we considered  the theory of a set of two
types of particles coupled to a BF topological term (in
$2+1$-dimensions). Although this theory has an interest on its
own, because it has a direct relationship with the problem of
interacting anyons, it also serves to prepare the scene for the
study carried out in the last section, where we consider a model
that involves extended objects (strings) and particles interacting
through a BF term in 3+1 dimensions. In both models, quantization
of the corresponding ``charge'' of the material objects involved
(i.e., point particles or strings) is necessary for the
consistence of the geometric representation.

Also, both models share the following feature: the topological
interaction can be casted into a kind of multivaluedness of the
corresponding wave functionals, that in the geometrical
representation is manifested through the functional dependence on
the winding number of a path around a point in the plane (in the
$2+1$ dimensions case), the winding number of a path around a
closed string, or the wrapping number of a surface around a point
(in the $3+1$ BF model).

To conclude, it is interesting to point out that, as in the $2+1$
model of anyons, it is possible to decouple the center of mass and
the relative motions in the particle-string model through the
introduction of the variables
\begin{eqnarray}\label{4.38}
\vec{r}_{rel}=\vec{r}-\vec{z}(\sigma); \qquad
\vec{R}_{CM}=\frac{m\vec{r}+\alpha\int\vec{z}(\sigma)d\sigma}{m+\alpha}.
\end{eqnarray}
Then, the lagrangian can be alternatively written down as
\begin{equation}\label{4.39}
L=\frac{1}{2}m\dot{\vec{R}}_{CM}^{2}+\frac{\alpha}{2}\int(
\dot{\vec{r}}_{rel}^{2}-\vec{r}^{'2}_{rel})d\sigma-\frac{\alpha^{2}}{2(m+\alpha)}
\left(\int\dot{\vec{r}}_{rel}^{2}d\sigma\right)^{2}-\frac{q\phi}{4\pi}\dot{\Omega}.
\end{equation}
Hence, in the particle-string model, the topological interaction
contributes to the lagrangian as the total derivative of a
multivalued function, namely, the solid angle subtended by the
string measured from the particle´s position. This feature is a
nice generalization of what occurs in its relative $2+1$
dimensional model of two anyons, and we believe that their
consequences deserve to be further considered \cite{Setaro}.

\section{Acknowledgments}
This work was supported by Project G-2001000712 of FONACIT. Also, the authors would 
like to thank the support given by OPSU.

\appendix

%\end{multicols}
\end{document}